\documentclass[structabstract]{aa}
\usepackage{txfonts}
\usepackage{appendix}
\usepackage{natbib}
\usepackage{multirow}
\usepackage{graphicx}
\usepackage{url}
\usepackage{lscape}
\bibpunct{(}{)}{;}{a}{}{,}

\begin{document}

\title{Variable X-ray absorption in the mini-BAL QSO PG~1126-041\thanks{Based
on observations obtained with XMM-\textit{Newton}, an ESA science mission
with instruments and contributions directly funded by ESA Member States and
NASA.}}

\author{
	M.~Giustini\inst{1,}\inst{2,}\inst{3,}\inst{7}
	\and
	M.~Cappi\inst{2}
	\and
	G.~Chartas\inst{4}
	\and
	M.~Dadina\inst{2}
	\and
	M.~Eracleous\inst{3,5}
	\and
	G.~Ponti\inst{6}
	\and
	D.~Proga\inst{7}
	\and
	F.~Tombesi\inst{8,}\inst{9}
	\and
	C.~Vignali\inst{1,}\inst{10}
	\and
	G.G.C.~Palumbo\inst{1}
	}

\offprints{M.~Giustini, \email{giustini@iasfbo.inaf.it}}

\institute{\inst{1} Dipartimento di Astronomia, Universit\`a degli Studi di Bologna, via Ranzani 1, I-40127, Bologna, Italy\\
           \inst{2} INAF-Istituto di Astrofisica Spaziale e Fisica cosmica di Bologna, via Gobetti 101, I-40129, Bologna, Italy \\
           \inst{3} Department of Astronomy and Astrophysics, the Pennsylvania State University, 525 Davey Lab, University Park, PA 166802, USA\\
\inst{4} Department of Physics and Astronomy, College of Charleston, Charleston, SC 29424, USA\\
           \inst{5} Center for Gravitational Wave Physics, the Pennsylvania State University, University Park, PA 166802, USA\\
           \inst{6} School of Physics and Astronomy, University of Southampton, Highfield, Southampton, SO17 1BJ, UK \\
           \inst{7} Department of Physics and Astronomy, University of Nevada Las Vegas, 4505 Maryland Pkwy Las Vegas, NV 891541-4002, USA\\
           \inst{8} X-ray Astrophysics Laboratory, NASA/Goddard Space Flight Center, Greenbelt, MD 20771, USA \\ 
           \inst{9} Department of Astronomy and CRESST, University of Maryland, College Park, MD 20742, USA \\
           \inst{10} INAF-Osservatorio Astronomico di Bologna, via Ranzani 1, I-40127, Bologna, Italy\\}

\date{Received XXXXXXXXXX XX, XXXX; accepted XXXXXXXXXX XX, XXXX}

\abstract {X-ray studies of active galactic nuclei (AGN) with powerful nuclear
winds are important for constraining the physics of the inner accretion/ejection
flow around supermassive black holes (SMBHs) and for understanding the impact of
such winds on the AGN environment.}{Our main scientific goal is to constrain the
properties of the circum-nuclear matter close to the SMBH in the mini-broad absorption line quasar (mini-BAL QSO) PG
1126-041 using a multi-epoch observational campaign with XMM-\textit{Newton}.}
{We performed temporally resolved X-ray spectroscopy and simultaneous UV and
X-ray photometry on the most complete set of observations and on the deepest
X-ray exposure of a mini-BAL QSO ever.} {We found complex X-ray spectral
variability on time scales of both months and hours, which is best reproduced by
means of variable massive ionized absorbers along the line of sight. As a
consequence, the observed optical-to-X-ray spectral index is found to be
variable with time. In the highest signal-to-noise observation we detected
highly ionized X-ray absorbing material outflowing much faster ($\upsilon_X\sim
16\,500$~km~s$^{-1}$) than the UV absorbing one ($\upsilon_{uv}\sim
5\,000$~km~s$^{-1}$). This highly ionized absorber is found to be variable on
very short (a few kiloseconds) time scales.}{Our findings are qualitatively
consistent with line-driven accretion disk winds scenarios. Our observations
have opened the time-resolved X-ray spectral analysis field for mini-BAL QSOs.
Only with future deep studies will we be able to map the dynamics of the inner
flow and understand the physics of AGN winds and their impact on the
environment.}


\titlerunning{Variable X-ray absorption in PG 1126-041}
\authorrunning{M.~Giustini et al.}

\maketitle

\section{Introduction}

The structure of the inner regions (sub-parsec scale) of active galactic nuclei
(AGN), as probed by UV and X-ray observations, seems to be very complex, and
certainly it is still not understood well. The optical/UV continuum emission is
most probably due to the thermal emission from an optically thick, geometrically
thin accretion disk \citep{1973A&A....24..337S}. The disk surrounds the central
supermassive black hole (SMBH), spanning radii from a few up to several hundreds
of gravitational radii ($r_g\equiv GM_{\rm{BH}}/c^2$), i.e. possibly from the
innermost stable circular orbit around the SMBH up to the disk
self-fragmentation radius \citep[e.g.,][]{1981ARA&A..19..137P}. The origin of
the X-ray continuum emission is less clearly understood and is thought to be the
result of the Comptonization of accretion disk UV seed photons into a ``cloud''
of very hot electrons, the so-called X-ray corona
\citep{1991ApJ...380L..51H,1994ApJ...432L..95H}. What is clear both from
variability and microlensing studies is that the X-ray emission region is much
smaller than the UV one, and it spans only a few up to tens of gravitational
radii
\citep[e.g.,][]{2008ApJ...689..755M,2009ApJ...693..174C,2010ApJ...709..278D}. An
extensive X-ray monitoring of the inner regions of the nearby Seyfert 1.8 NGC
1365 has made it possible to constrain the size of the X-ray emitting region to
again be a few $r_g$, using a totally independent technique
\citep{2009MNRAS.393L...1R}.
The UV and X-ray continuum photons are observed to be reprocessed (absorbed,
re-emitted, scattered, reflected) by gas and dust in the inner regions of AGN,
so that we almost never observe just the primary continuum emission of these
objects \citep[see e.g. the comprehensive analysis of ][again on NGC 1365]
{2005ApJ...623L..93R,2005ApJ...630L.129R,2007ApJ...659L.111R,
2009ApJ...696..160R}. Conversely, observing the reprocessing features in the UV
and X-ray bands can put strong contraints on the geometry, the physical
characteristics, and the dynamics of the gas hosted in the inner regions of AGN.
In the past few decades, spectroscopy in the UV and X-ray band have revealed the
presence of substantial column densities of ionized gas \textit{flowing out}
from the inner regions of AGN. 

In the UV band we observe blueshifted spectral
absorption lines due to resonant transitions of ionized metals such as Mg~II,
Al~III, Si~IV, C~IV, N~V, O~VI. Depending on the width of the absorption
troughs, quasars hosting such features are classified as broad absorption line
quasars \citep[BAL QSOs, FWHM $> 2000$~km~s$^{-1}$, e.g.][]{1980ApJ...238..488T,
1991ApJ...373...23W, 2003AJ....126.2594R}, mini-broad absorption line quasars
(mini-BAL QSOs, $500$~km~s$^{-1}<$ FWHM $<2000$~km~s$^{-1}$), and narrow
absorption line quasars \citep[NAL QSOs, FWHM $<500$~km~s$^{-1}$;
see][]{2001ApJ...549..133G,2004ASPC..311..203H}. BALs features are observed in
$\sim 15\%$ of optically selected QSOs
\citep[e.g.,][]{2003AJ....125.1784H,2003AJ....126.2594R,
2008MNRAS.386.1426K,2011MNRAS.410..860A}. Mini-BALs and NALs are together
observed in $\sim 12-30\%$ of optically selected QSOs
\citep{2008ApJ...672..102G}. Despite the very different absorption trough widths
observed, BAL, mini-BAL, and NAL QSOs share the same velocity ranges, reaching
UV terminal velocities $\upsilon_{out}^{UV}$ from a few 10$^2$~km~s$^{-1}$ up to
several 10$^4$~km~s$^{-1}$ \citep[e.g.,][]{2007ASPC..373..287R,
2007ApJS..171....1M, 2009ApJ...692..758G}. Furthermore, $\upsilon_{out}^{UV}$ is
known to correlate with the continuum luminosity $L_{UV}$
\citep{2002ApJ...569..641L, 2008ApJ...672..102G}, meaning that the highest
terminal velocities are observed in the objects with the highest UV continuum
luminosity, suggesting an important role of the AGN UV radiation pressure in the
acceleration of such winds.

In the X-ray band absorption due to ionized species, such as N~VI-VII,
O~VII-VIII, Mg XI-XII, Al XII-XIII, Si~XIII-XVI, as well as L-shell transitions
of Fe XVII-XXIV, are observed to be blueshifted by a few hundred to a few
thousand km~s$^{-1}$ in $\sim 50\%$ of type 1 AGN \citep[the ``warm absorber'',
e.g. ][]{1997MNRAS.286..513R, 2005A&A...432...15P, 2007MNRAS.379.1359M}.
Moreover, in recent years, thanks to the high collecting area of X-ray
satellites such as XMM-\textit{Newton}, \textit{Chandra}, and \textit{Suzaku},
blueshifted absorption lines due to highly ionized gas (i.e. Fe~XXV, Fe~XXVI)
outflowing at much higher velocity ($\upsilon_{out}\sim$0.05-0.2$c$) have been
observed in a number of AGN, at both low and high redshift \citep[see e.g.][and
references therein]{2003MNRAS.345..705P, 
2006AN....327.1012C, 2010A&A...521A..57T,
2010ApJ...719..700T}. In particular, X-ray BALs in He-like iron, blueshifted up
to $\upsilon_{out}=0.7c$, have been observed in gravitationally lensed BAL and
mini-BAL QSOs \citep{2002ApJ...579..169C, 2003ApJ...595...85C,
2009ApJ...706..644C}.

Theoretically, powerful winds can be launched from AGN accretion disk either by
thermal, magnetic, or radiative pressure. In either case, a wind will be
launched only if the pressure overcomes the SMBH gravity pull. As a rule of
thumb, the closer to the SMBH the launching point is, the higher the wind
terminal velocity. \citep[see e.g.][for a
review]{2006MmSAI..77..598K,2007Ap&SS.311..269E,2007ASPC..373..267P}.
Observations of the wind outflow velocity are then crucial in order to constrain
the launching radius, and hence the physical driving mechanism. The thermal
mechanism can account for low-velocity outflows arising at large radii
\citep[e.g. the X-ray warm absorber,][]{2005ApJ...625...95C}. Instead, in order
to explain the extreme velocities observed in the other absorbers, either
magnetic or radiation pressure must be invoked. For instance, line driven
accretion disk wind models show that at high accretion rates, the UV radiation
pressure from the disk can be high enough to launch powerful winds from
distances of a few tens of $r_g$. In the inner part of this flow,
high column densities of X-ray absorbing gas shield the wind from the strong
X-ray continuum source, preventing it from becoming overionized
\citep{1995ApJ...451..498M, 2000ApJ...543..686P, 2004ApJ...616..688P}. Thus, the
geometry and the dynamics of such winds depend critically on the UV/X-ray flux
ratio and on the AGN accretion rate. On the other hand, for the launch and
acceleration of magnetically driven disk winds, there is no need for X-ray
shielding, and the accretion rate is not a critical parameter; however, the ionization
state of the gas, which is a critical parameter, will still depend on
them \citep[e.g.,][]{1994ApJ...434..446K,2005ApJ...631..689E}.

BAL and mini-BAL QSOs are actually known to be ``X-ray weak'' with respect to
the average QSOs. A comparison of the X-ray fluxes of BAL and non-BAL QSOs with the
same UV flux indicates that BAL QSOs typically have X-ray fluxes that are
$10$ to $30$ times lower than those of non-BAL QSOs \citep{1995ApJ...450...51G,
1997ApJ...477...93L}. A number of studies \citep[e.g.][]{2000ApJ...528..637B,
2001ApJ...558..109G, 2006ApJ...644..709G} strongly support absorption as the
reason for the X-ray weakness. Recently a number of ``X-ray bright''
BAL and mini-BAL QSOs have been discovered, with the latter class X-ray brighter and
less X-ray absorbed than the former \citep{2008A&A...491..425G,
2009ApJ...696..924G}. The NAL QSOs have been shown to be generally unabsorbed in
X-rays, and ``X-ray brighter'' than both mini-BAL and BAL QSOs
\citep{2009NewAR..53..128C}.

The physical link between the UV and X-ray outflowing absorbers and the
dynamical behavior of the wind are far from being understood. Reported
estimates of the distances of the absorbers from the central SMBH range from the
inner regions of the accretion disk \citep[e.g.,][]{2009ApJ...706..644C} to the
parsec-scale torus \citep[e.g.,][]{2005A&A...431..111B} to the kiloparsec
scales \citep[e.g.,][]{2010ApJ...709..611D}. Currently the uncertainties of the
launching radii are relatively large, which translates to a high uncertainty on
the mass outflow rate and on the kinetic energy injection associated with such
winds. It could also be that the launching radii depend on the mass of the
SMBH. Constraining the physical mechanism responsible for launching and
accelerating AGN winds, i.e. understanding the link between the accretion and
ejection processes in AGN, can help not only in understanding the evolution of
these sources, but also in quantifying their impact on the surrounding
environment, i.e. the amount of feedback \citep[see e.g.][]{2005Natur.433..604D,
2005ApJ...635L..13S, 2006MmSAI..77..573E}.

In this article we present the results of an XMM-\textit{Newton} observational
campaign on the mini-BAL QSO PG 1126-041, which provides both the largest
dataset (four pointings) and the deepest X-ray exposure (130 ks) on a mini-BAL
QSO to date. The source is extremey interesting for showing all three
absorbers discussed above: mini-BALs in the UV, an X-ray warm absorber, and a
highly ionized, high-velocity X-ray outflow. Constraining the X-ray/UV
properties of this source can give precious insight into the physics of AGN
winds.

The structure of the article is as follows. Section~\ref{PG} summarizes the
source properties known from the literature. Section~\ref{OBS} reports on the new
XMM-\textit{Newton} observations and data reduction. EPIC pn timing analysis
results are first presented in Section~\ref{timing}. EPIC pn and MOS spectral
analysis is then presented in Sections~\ref{2009average} (average spectra of each
epoch) and \ref{timres} (time-resolved 2009 Long Look pn spectra). Section~\ref{OM}
contains the results of simultaneous optical/X-ray photometric analysis. Results
are discussed in Section~\ref{DISC} and conclusions presented in
Section~\ref{CONC}.

A cosmology with $H_0=70\:$km~s${-1}$~Mpc$^{-1}$, $\Omega_{\Lambda}=0.73$ and
$\Omega_{\rm{M}}=0.27$ \citep{2011ApJS..192...18K} is used throughout the paper.

\section{PG 1126-041}\label{PG} PG~1126-041 (a.k.a. Mrk 1298) is a nearby
($z=0.06$), radio-quiet AGN with a luminosity in between those that are typical
of Seyferts and QSOs \citep[$M_B=-22.8$, $L_{\rm{bol}}\sim
10^{12}L_{\odot}$,][]{1983ApJ...269..352S,1989ApJ...347...29S,2001AJ....122.2791S}.
Its optical strong Fe~II and weak [O~III] line emission
\citep{1992ApJS...80..109B} are characteristics of narrow line Seyfert 1
galaxies (NLS1), although the FWHM of its H$\beta$ line of 2150 km s$^{-1}$ is
slightly more than the 2000 km s$^{-1}$ value used in defining NLS1s
\citep{1981ApJ...250...55S}. 
Its black hole mass is estimated to be $\log
(M_{\rm{BH}}/M_{\odot})\sim 7.7$ using the $L_{\lambda}(5100\AA)$ continuum
luminosity density and the FWHM(H$\beta$) measured from a single-epoch optical
spectrum and the empirical relations reported by \citet{2006ApJ...641..689V}.
However, the stellar velocity dispersion in the host galaxy of PG 1126-041
measured from the stellar CO absorption line width in the VLT H-band spectra
gives a slightly higher black hole mass of $\log (M_{\rm{BH}}/M_{\odot})\sim
8.1$ \citep{2007ApJ...657..102D}. 

PG~1126-041 is also classified as ``soft X-ray
weak''\footnote{Soft X-ray weak sources are those being characterized by an
observed optical-to-X-ray spectral index $\alpha_{\rm{ox}}< -2$, see
Section~\ref{OM}.} by \citet{2000ApJ...528..637B}, who measured an
$\alpha_{\rm{ox}}\sim -2.19$ based on \textit{ROSAT} X-ray observations and on
the optical flux density as given by \citet{1987ApJS...63..615N}.
\citet{1986ApJ...305...57T} measured an $\alpha_{\rm{ox}}=-1.90$ referring to an
earlier \textit{Einstein} X-ray observation and the $2500\,\AA$ rest frame flux
density as given by \citet{1983ApJ...269..352S}. Neither of these
$\alpha_{\rm{ox}}$ measurements were based on simultaneous X-ray and optical/UV
observations. 
\citet{1999MNRAS.307..821W} report on the \textit{IUE} and
\textit{ROSAT} observations of PG~1126-041. The 1992 \textit{IUE} observation
shows a high-velocity UV absorbing outflow as mini-BALs in the C~IV, N~V, and
Si~IV ions blueshifted up to $\sim 5\,000$~km~s$^{-1}$. The simultaneous
\textit{ROSAT} pointing revealed a 0.2-2~keV spectrum best-fitted by a steep
powerlaw emission with $\Gamma\sim 2.8$, absorbed by a column density $N_{\rm
W}\sim 3\times 10^{22}$~cm$^{-2}$ of ionized gas, $\log\xi\sim
1.7$~erg~cm~s$^{-1}$. A subsequent \textit{IUE} spectrum taken in 1995 showed a
continuum that was brighter by a factor of two and weaker absorption lines.
\citet{2000A&A...354..411K} confirm the detection of ionized absorption in the
1992 \textit{ROSAT} spectrum and also suggest X-ray temporal variability on
timescales as short as 800~s.
The main properties of PG 1126-041 are
summarized in Table~\ref{properties}.

\begin{table}
\caption{Main properties of PG 1126-041}
\label{properties}
\centering          
\begin{tabular}{l l l}  
\hline
\hline
RA &  11h 29m 16.6s  & (a) \\
DEC &  -04h 24m 08s & (a) \\
$z$ & 0.062 & (a) \\
 $D_L$ & 284 Mpc & \\
 $N_H^{gal}$ & $4.35\times 10^{20}$ cm$^{-2}$ & (b) \\
 $E(B-V)$ &  0.055  mag  & (c)   \\
 $M_{\rm{BH}}$   &  $5.0-12\times 10^7 M_{\odot}$ & (d-e) \\
  m$_B$ & 14.92  mag & (a) \\
  $L_{bol}$ & $8.9\times 10^{11}L{\odot}$& (f)\\
  $L_{2-10}$ & $2\times 10^{43}$ erg s$^{-1}$ & (h) \\
$  \upsilon_{\rm out}^{\rm uv}$ & 5000 km s$^{-1}$ & (g)	\\
$  \upsilon_{\rm out}^{\rm X}$ & 16500 km s$^{-1}$ & (h)	\\
\hline
\end{tabular}\\
\medskip 
 \flushleft{References: (a) \citet{1983ApJ...269..352S}; (b)
 \citet{2005A&A...440..775K}; (c) \citet{1998ApJ...500..525S}; (d) \citet{2006ApJ...641..689V}; (e)
 \citet{2007ApJ...657..102D}; (f) \citet{2001AJ....122.2791S}; (g)
 \citet{1999MNRAS.307..821W}; (h) This work, referred to the 2009 Long Look observation.}
 \end{table}

\section{Observations and initial data reduction}\label{OBS}

\begin{table*}
\caption{X--ray observation log}
\label{table:obslog}
\centering          
\tiny{\begin{tabular}{c c c c c c c c c }  
\hline\hline
  OBSID & Date & Duration & Exposure & Mode & Filter & Count-rate & $\log f_{0.2-10}$  & Epoch \\
            &              &               & M1 / M2 / pn &   &    &  MOS / pn  & pn & \\
            &                &[ks]    & [ks]     &  & & [0.1 ct s$^{-1}$] &  [erg cm$^{-2}$ s$^{-1}$] &\\
(1)      & (2)      & (3)     & (4)           & (5)            & (6)       & (7)    & (8)      & (9) \\
\hline
 & & & & & & & & \\
 0202060201 & 12/31/2004 & 33.8 & 32.9 / 32.9 / 28.3 & LW & t& 0.61$\pm {0.02}$ / 2.01$\pm {0.03}$ & -11.88  &  Dec. 2004\\
             0556230701 & 06/15/2008 & 31.4 & 3.7 / 3.5 / 3.2    & FF  & m&  0.91$\pm {0.05}$  / 3.01$\pm {0.10}$&  -11.70   & Jun. 2008\\
             0556231201 & 12/13/2008 & 11.9 & 4.9 / 4.7 / 3.7    & FF  & m & 1.68$\pm {0.06}$ / 5.93$\pm {0.20}$& -11.52  & Dec. 2008 \\
             0606150101 & 06/21/2009 & 134.3& 85.4 / 89.2 / 91.2 & FF & m & 0.57$\pm {0.01}$ / 1.96$\pm {0.02}$ &  -11.93 & 2009 Long Look\\
 & & & & & & & & \\
\hline
\end{tabular}}\\
\begin{flushleft}
\tablefoot{Col.(1): Observation ID; Col.(2): Date of observation; Col.(3):
Nominal duration of the observation; Col.(4): Net exposure time for each
instrument after the background flaring filtering was applied; Col.(5):
Observing Mode, LW= Large Window, FF=Full Frame; Col.(6): Optical Filter
applied, t=thin, m=medium; Col.(7): Net count-rate in the 0.2-10~keV and
0.3-8~keV range for the EPIC-pn and EPIC-MOS instruments, respectively, after
the local background subtraction; Col.(8): Observed EPIC-pn flux in the 0.2-10 keV
band;  Col.(9): Epoch name as used in the
text.}
\end{flushleft}
\end{table*}

XMM-\textit{Newton} observed PG~1126-041 four times, once in 2004 December,
twice in 2008 June and December, and once more in 2009 June. In Table
\ref{table:obslog} we list the main parameters of each dataset. We reduced all
the datasets using the XMM-\textit{Newton} SAS v.10.0.0 and calibration files
generated in August 2010. For each dataset we extracted the light curve from the
whole field of view of events with energies greater than 10 keV for each European
Photon Imaging Camera (EPIC) MOS cameras, and with energies 10 keV $< E <$ 12
keV for the EPIC-pn instrument. We then filtered out the time periods in which
there was significant background flaring. Source counts from the cleaned event
files were extracted from circular regions centered on the PG 1126-041 position,
with $\sim$30--45'' radii, depending on the signal-to-noise ratio (S/N) as
determined with the \texttt{eregionanalyse} task. Background counts were
extracted from circular regions of the same area as the source ones. We retained
up to double events (flagged as $\# XMMEA\_EP$) for the pn dataset and quadruple
events (flagged as $\# XMMEA\_EM$) for the MOS datasets. For each dataset we
generated the ancillary response file and the redistribution matrix file at the
source position using the tasks \texttt{arfgen} and \texttt{rmfgen}. Source
light curves were created in different energy bands and corrected with the
\texttt{epiclccorr} task for both the local background and the mirrors/detector
inefficiencies. Due to the low X-ray fluxes involved, pile-up effects are
negligible in our spectral analysis. For the same reason we cannot make use of
the reflection grating spectrometer (RGS) aboard XMM-\textit{Newton}. Optical
monitor (OM) data were all taken in Image mode and were processed using the
standard procedure as suggested by the XMM-\textit{Newton} SOC.

\section{Timing analysis}\label{timing}
\begin{figure*}
\centering
\includegraphics[width=18cm]{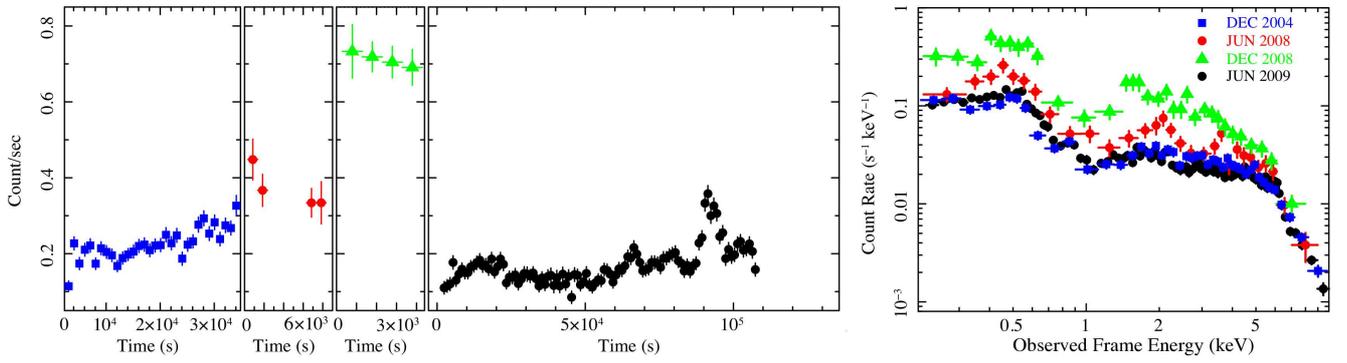}
\caption{Variability of PG 1126-041 on month time scales. Left panel: background subtracted
0.2-10 keV pn light curves extracted in the four different epochs of XMM-\textit{Newton}
observations and binned to 1 ks: December 2004 (blue squares),
June 2008 (red circles), December 2008 (green triangles), and 2009 Long Look
(black circles). Right panel: the time-averaged pn spectra corresponding to the
time intervals shown on the left and using the same color
codes.\label{FIG1}}
\end{figure*}
\begin{figure*} 
\includegraphics[width=18cm]{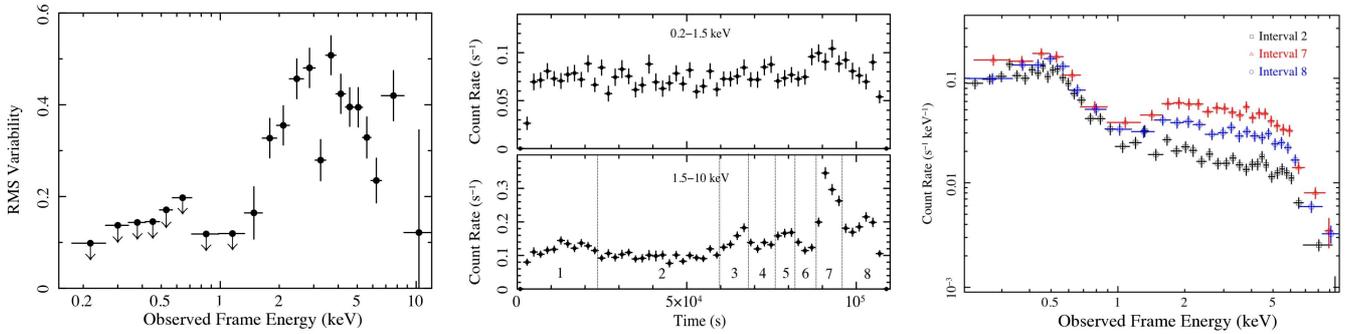} 
\caption{Variability on kilosecond time scales detected during the 2009 Long Look observation. 
Left panel: RMS variability computed for the pn observation over the
0.2-10 keV band; arrows mark upper limits, error bars are at 1$\sigma$ confidence level. 
Middle panel: 0.2-1.5 and 1.5-10 keV background-subtracted pn light curves binned to 2 ks; the numbers in the 1.5-10
keV light curve mark the eight time intervals on which time-resolved spectral
analysis was performed. 
Right Panel: three representative spectra extracted during the 2nd, the 7th, and the 8th time intervals of the
2009 Long Look observation. \label{FIG2}} 
\end{figure*} 

The lefthand panel of Fig.~\ref{FIG1} shows the background-subtracted 0.2-10~keV
pn light curves of PG 1126-041 for each epoch of observation, binned to 1 ks.
Compared to the light travel time $t_L=r_g/c$, the time bin corresponds to $\sim
1.5-4 r_g$ for the two different estimates for the black hole mass of PG
1126-041.
The variations on month timescales are dramatic: the source flux increases by a
factor of $\sim 4$ from December 2004 to December 2008, then
comes back to a low count-rate regime in June 2009. For comparison, the average
0.2-10~keV pn spectra of each epoch of observation are shown in the righthand panel
of Fig.~\ref{FIG1}. Most of the spectral variability on month time scales
occurs at energies $E\lesssim 6$~keV. The spectral shape is peculiar and
obviously deviates from a simple power-law. A prominent broad absorption
 feature is evident in all the spectra around $E\sim$0.6-1.5~keV.

The high S/N ratio and the long contiguous good exposure time allowed us to
perform a more detailed timing analysis on the 2009 Long Look pn exposure.
Figure~\ref{FIG2} summarizes the results. The lefthand panel shows the root mean
square variability \citep[RMS; ][]{2003MNRAS.345.1271V, 2004A&A...417..451P}
computed on the 0.2-10 keV band: on very short time scales (ks), the source flux
varies up to 50\% at energies $E\gtrsim 1.5$~keV, while it is constant at
lower energies. The middle panel shows the background-subtracted lightcurves
extracted in the 0.2-1.5 and 1.5-10 keV bands; following the 1.5-10 keV band flux
fluctuations, we split the exposure in eight time intervals on which we
perform temporally resolved spectral analysis. Three representative spectra
extracted during the 2nd, the 7th, and the 8th time intervals are shown in the
righthand panel, and most of the spectral variability on ks time scales occurs at $1.5$
keV $\lesssim E \lesssim 9$ keV.

From the timing analysis, it follows that the X-ray spectra of PG 1126-041 can
be divided into two spectral components: one dominates at $E\gtrsim 1.5$~keV and
contributes 100\% to the ks time scale variability, while the other one
dominates at lower energies and is significantly variable only on longer
(month) time scales. With these considerations in mind, we proceed with the
analysis of both the average (Sect.~\ref{2009average}) and the time-resolved
(Sect.~\ref{timres}) spectra.

\section{Spectral analysis: the average spectra}\label{2009average}
In our spectral analysis we used the \texttt{Xspec v.12.6.0} software. 
All the spectra were grouped to a minimum number of 20 counts per energy bin 
in order to apply the $\chi^2$
statistics in the search for the best-fit model \citep{1976ApJ...210..642A}.
Errors are quoted at 1$\sigma$ confidence, unless otherwise stated . All the
models include Galactic absorption by a column density $N_{\rm
H}^{\rm{Gal}}=4.35\times 10^{20}$~cm$^{-2}$ \citep{2005A&A...440..775K}. We
focused our analysis on the 0.2-10 keV band of the pn data and on the 0.3-8 keV
band of the MOS data.
We simultaneously fit the average spectra of the four epochs, 
Dec. 2004, Jun. 2008, Dec. 2008, and 2009 Long Look, to investigate
the causes of the observed variability on month-long time scales. 
For each epoch, the model parameters were kept fixed between the MOS and
the pn datasets, except for the primary power-law normalizations that
were left free to vary within 10\% to account for the instrumental cross-calibration
uncertainties. For the sake of clarity, in most of the following plot we only show the 
pn data.

 \begin{figure}
\includegraphics[width=84mm]{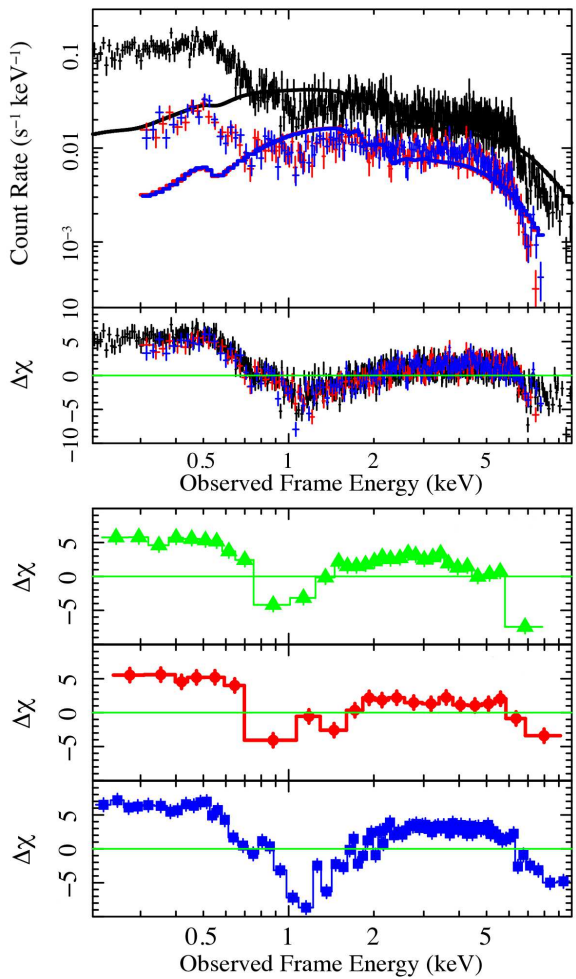}
\caption{Top panel: average 2009 Long Look pn (black), MOS1 (red), and MOS2 (blue) spectra
modeled with a power-law emission absorbed by the Galactic interstellar medium, along
with the residuals to the model, scaled by the error bar. Bottom panel:
spectral residuals to the power-law model for the other three epochs:
Dec. 2008 (green triangles), Jun. 2008 (red circles), 
and Dec. 2004 (blue squares); the data have been rebinned
for visual purpose only to respectively 5, 5, and 7$\sigma$.\label{FIG3}}
\end{figure}	

The spectra were initially fit to a simple power-law model, with fixed slope and
free amplitude between different epochs. Results are shown in Fig.~\ref{FIG3}.
The fit is very poor, with a chi square per degrees of freedom
$\chi^2/\nu=8801/1591$, and a very flat photon index $\Gamma\sim 0.6$. The
spectral residuals of all the four epochs have a similar shape, and indeed
allowing for the spectral slope of the power-law to vary between epochs does not
improve the fit statistics. Three main deviations from the simple power-law
model are consistently observed in spectra from all three instruments and in
all four epochs of observation: (a) both negative and positive residuals in
the soft band, $E\lesssim 1.5$~keV, (b) a deficit of counts at the highest
energies probed by the EPIC camera, $E\sim 7-10$~keV, and (c) a cutoff at
$E\lesssim 2$~keV.

We then tested several models of increasing complexity, to reproduce
the complex spectral shape of PG 1126-041. A neutral absorber fully covering the
emission source is not required by the data. On the other hand, allowing for
the absorber to only partially cover the source (\texttt{zpcfabs} model in
\texttt{Xspec}) significantly improves the fit statistics with respect to the power-law
model ($\Delta\chi^2/\Delta\nu=5148/8$), but still gives an unacceptable fit
($\chi^2_r\sim 2.3$). A better fit for the same number of degrees of freedom
($\Delta\chi^2/\Delta\nu=6567/8$) is given by a ionized absorber totally covering the
source (modeled with \texttt{Xstar}\footnote{ We applied the precompiled
\texttt{grid25}, publicly available at
\url{ftp://legacy.gsfc.nasa.gov/software/plasma_codes/xstar/xspectables/grid25/},
which is computed for an ionizing continuum with $\Gamma=2$, a gas shell with
$n=10^{12}$~cm$^{-3}$, a turbulent velocity $\upsilon_{turb}=200$~km~s$^{-1}$,
and solar abundances, see \citet{2001ApJS..133..221K} and the \texttt{Xstar}
documentation.}), but again, the fit is statistically unacceptable ($\chi^2_r\sim
1.5$).

The ionized absorber was then allowed to only partially cover the X-ray
source (model A), by adding a secondary soft power-law with the same slope as
the primary one. The ratio between the normalizations of the two power-laws
gives the fraction of the direct (unabsorbed by intrinsic absorption) flux $F$; the absorber covering factor is then
$C_f = \left(1-F\right)$. We obtain an improvement in the fit statistics of
$\Delta\chi^2/\Delta\nu=341/4$, and the resulting fit is marginally acceptable,
with a $\chi^2_r\sim 1.199$. 
The model A parameters are reported in Table~\ref{TABLEFITAVERAGE} along with 
their 1$\sigma$ statistical errors. 
We note how the direct flux fraction in every epoch is around 2-3\%, as is observed for the
scattered component in absorbed AGN, and scattered emission accompanied by
recombination emission lines would be naturally associated to
the presence of ionized gas along the line of sight. 
Unfortunately, with the S/N of our observations we cannot distinguish among the 
partial covering and scattering scenarios in reproducing the soft band residuals,
and the secondary soft power-law physical interpretation
is not unambiguous.
To the first order, model A is able
to reproduce the complex spectral shape of PG 1126-041, in particular
by recovering the cutoff at $E\lesssim 2$ keV and most of the soft band spectral
residuals. While the Jun. 2008 and Dec. 2008 spectra are generally
reproduced well, the low flux states of the 2009 Long Look and Dec. 2004 still show
strong residuals at $E\lesssim 1$ keV and $E\gtrsim 6$ keV.

\begin{table}
\caption{Spectral analysis results of the simultaneous fit to the average spectra of each epoch
\label{TABLEFITAVERAGE}}
\centering          
\small{\begin{tabular}{c c c c c}  
\hline\hline
                     & Jun. 2009 & Dec. 2008 & Jun. 2008 & Dec. 2004 \\[+0.05in]
                     \hline
 \multicolumn{5}{c}{\textit{Model A: ionized partial covering}}\\[+0.05in]
 \multicolumn{5}{c}{\texttt{Xspec: zpo$_2$ + XSTAR*zpo$_1$}}\\[+0.05in]
 $\Gamma$  & \multicolumn{4}{c}{2.15$^{+0.01}_{-0.01}$} \\[+0.05in]
 $N_{\rm{1\,keV}}$ & 1.05$^{+0.15}_{-0.03}$ & 1.89$^{+0.09}_{-0.08}$ & 1.40$^{+0.06}_{-0.06}$ & 1.00$^{+0.02}_{-0.02}$ \\[+0.05in]
 $N_W^{m.i.}$ & 14.36$^{+0.61}_{-0.47}$ & 4.91$^{+0.24}_{-0.25}$ & 7.97$^{+0.36}_{-0.24}$ & 9.78$^{+0.18}_{-0.08}$ \\[+0.05in]
 $\log\xi^{m.i.}$ & 1.66$^{+0.02}_{-0.02}$ & 1.44$^{+0.06}_{-0.05}$ & 1.58$^{+0.02}_{-0.02}$ & 1.57$^{+0.01}_{-0.01}$ \\[+0.05in]
 $C_f^{m.i.}$ & 0.97$^{+0.01}_{-0.01}$ & 0.98$^{+0.01}_{-0.01}$ & 0.98$^{+0.01}_{-0.01}$ & 0.97$^{+0.01}_{-0.01}$ \\[+0.05in]
 $\chi^2/\nu$ & \multicolumn{4}{c}{1893/1579} \\[+0.05in]
 \hline
 \multicolumn{5}{c}{\textit{Model B: ionized partial covering + highly ionized absorber}}\\[+0.05in]
  \multicolumn{5}{c}{\texttt{Xspec: zpo$_2$ + XSTAR1*XSTAR2*zpo$_1$}}\\[+0.05in]
 $\Gamma$  & \multicolumn{4}{c}{2.14 $^{+0.01}_{-0.03}$} \\[+0.05in]
 $N_{\rm{1\,keV}}$ & 1.12$^{+0.06}_{-0.13}$ & 1.82$^{+0.11}_{-0.10}$ & 1.41$^{+0.09}_{-0.08}$ & 1.01$^{+0.03}_{-0.04}$ \\[+0.05in]
 $N_W^{m.i.}$ & 14.82$^{+0.52}_{-0.71}$ & 3.17$^{+0.24}_{-0.12}$ & 6.93$^{+0.57}_{-1.00}$ & 9.75$^{+0.24}_{-0.10}$ \\[+0.05in]
 $\log\xi^{m.i.}$ & 1.66$^{+0.01}_{-0.01}$ & 1.29$^{+0.04}_{-0.04}$ & 1.54$^{+0.02}_{-0.04}$ & 1.57$^{+0.01}_{-0.01}$ \\[+0.05in]
 $C_f^{m.i.}$ & 0.97$^{+0.01}_{-0.01}$ & 0.98$^{+0.01}_{-0.01}$ & 0.98$^{+0.01}_{-0.01}$ & 0.97$^{+0.01}_{-0.01}$ \\[+0.05in]
 $N_W^{h.i.}$ & 75.18$^{+0.55}_{-0.96}$ & $<6.82$ & $<6.15$ & 62.54$^{+27.45}_{-35.44}$ \\[+0.05in]
 $\log\xi^{h.i.}$ & 3.43$^{+0.09}_{-0.08}$ & $<3.00$ & $<2.73$ & 3.75$^{+0.12}_{-0.20}$ \\[+0.05in]
 $\upsilon_{out}^{h.i.}$ & 0.054$^{+0.09}_{-0.03}$& 0.0$^{\rm{F}}$ &  0.0$^{\rm{F}}$ &  0.0$^{\rm{F}}$ \\[+0.05in]
 $\chi^2/\nu$ & \multicolumn{4}{c}{1771/1570} \\[+0.05in]
 \hline
\end{tabular}}\\
\begin{flushleft}
\tablefoot{Units are the following: [$10^{-3}$ ph s$^{-1}$ keV$^{-1}$ cm$^{-2}$] for 
the power-law normalizations $N_{\rm{1\,keV}}$; [$10^{22}$ cm$^{-2}$] for the column densities
$N_W, N_W^{m.i.}$, and $N_W^{h.i.}$;  [erg cm s$^{-1}$] for the ionization parameters $\xi, \xi^{m.i.}$,
and $\xi^{h.i.}$; $c$ for the highly ionized absorber outflow velocity $\upsilon_{out}$;
the photon index $\Gamma$ and the covering fraction $C_f$
are adimensional parameters. Errors quoted are at 1$\sigma$ confidence level. a superscripted F denotes a fixed parameter.}
\end{flushleft}
\end{table}
\begin{figure}
\centering \includegraphics[width=80mm]{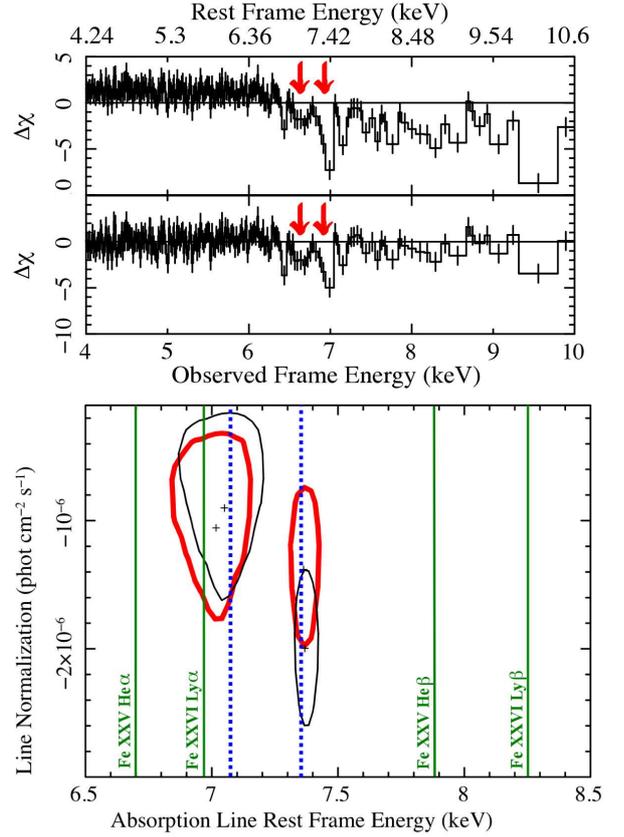} \caption{\label{FIG5} Top
panel: 4-10 keV spectral residuals of the 2009 Long Look pn data, relative to
the simple power-law model (top panel) and to model A (bottom panel). The two
arrows mark strong negative residuals that are present in both cases. Data have
been rebinned to 3$\sigma$ significance. Bottom panel: 99\% confidence contours
for the rest frame normalizations and centroid energy of the absorption lines
used to model the residuals; the black contours refer to the addition of the
absorption lines to the simple power-law model, while the red ones to the
addition to model A. The four vertical green lines mark the theoretical energies of
the strongest transitions of Fe XXV and Fe XXVI, while the two dotted blue lines
mark the position of the Fe XXV He$\alpha$ and Fe XXVI Ly$\alpha$ transitions,
both blueshifted by $\upsilon_{out}=0.055c$.} 
\end{figure} 
In particular, the top panel of Fig.~\ref{FIG5} shows a magnified view of the
4-10 keV residuals of the 2009 Long Look pn data, relative to the simple
power-law emission (top panel) and to model A (lower panel). We make an initial,
phenomenological model for these residuals by adding Gaussian profiles with
negative normalization to model A of the 2009 Long Look dataset. Even though the
model adopted is only a rough approximation of a more complex physical context,
the basic properties such as the FWHM and the EW of the absorption features are
still physically meaningful. Including two narrow (unresolved,
$\sigma\equiv 0$ eV) Gaussian absorption lines improves the fit statistics by
$\Delta\chi^2/\Delta\nu=60/4$ with respect to model A. The energies and
equivalent widths of the two absorption lines are
$E_1^{abs}=7.02^{+0.05}_{-0.02}$ and $E_2^{abs}=7.36^{+0.02}_{-0.02}$~keV,
$EW_1^{abs}=-80^{+20}_{-19}$ and $EW_2^{abs}=-127^{+24}_{-26}$~eV, in the source
rest frame. Allowing the widths of the absorption lines to be free parameters
does not improve the fit ($\Delta\chi^2/\Delta \nu=2/2$). 
In the bottom panel of Fig.~\ref{FIG5} we show the 99\% confidence contours for
the absorption lines centroid energy and (negative) normalization, 
along with the energies of the strongest transitions of Fe XXV and Fe XXVI.
There is a remarkable correspondence between the detected centroid energies
and the Fe XXV He$\alpha$ and Fe XXVI Ly$\alpha$ transitions,
both blueshifted by $\upsilon_{out}=0.055c$, 
so we identify the two absorption lines with these two transitions.

\begin{figure} \centering
\includegraphics[width=84mm]{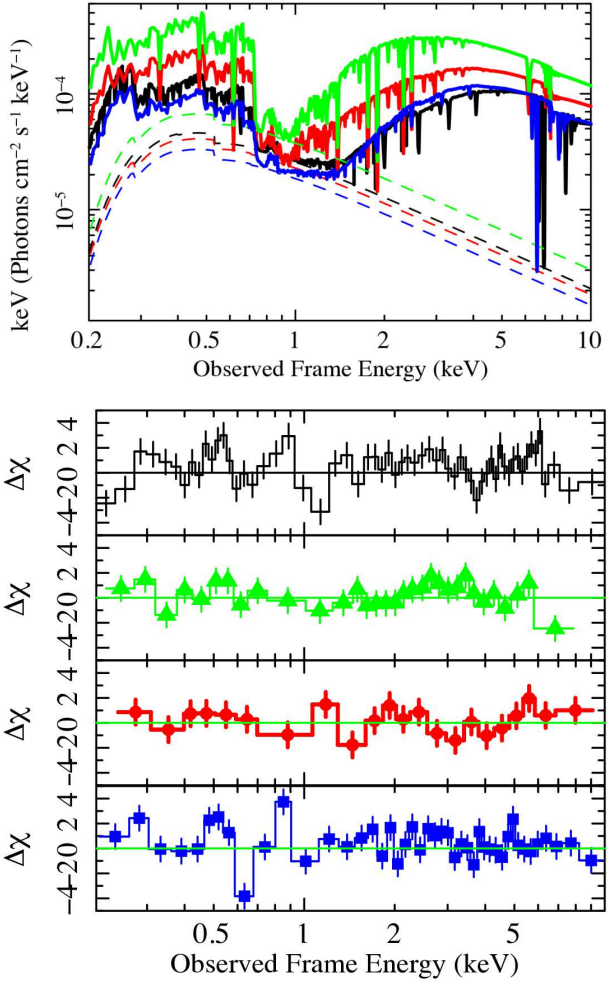} \caption{Top panel: the solid
lines show the power-law emission after the passage through two absorbers, one
of which is partially covering the source (model B). This model was fit to the
average 2009 Long Look (black), Dec. 2008 (green), Jun. 2008 (red), and Dec.
2004 (blue) epoch spectra. The dashed lines represent the direct continuum
components that are not absorbed. Bottom panel: average 2009 Long Look (black
points), Dec. 2008 (green triangles), Jun. 2008 (red circles), and Dec. 2004
(blue squares) spectral residuals relative to model B; the data have been
rebinned for visual purpose only to respectively 15, 5, 5, and
10$\sigma$.\label{FIG6}} \end{figure}

The moderately ionized absorber (m.i.) responsible for the opacity in the soft
band has an ionization state too low to account for the strong absorption in the
iron K band. So we added another layer of highly ionized (h.i.) gas modeled
again with \texttt{XSTAR}. We used the same input parameters as the \texttt{grid
25}, but we find a turbulence velocity $\upsilon_{turb}= 1500$~km~s$^{-1}$ to
better reproduce the residuals. Only in the case of the 2009 Long Look data we
allowed the velocity shift of the highly ionized absorber to be a free
parameter; given the much lower S/N of the three other epochs, we fixed the
velocity shift of this component to zero. The addition of the highly ionized
absorber significantly improves the fit with respect to the moderately ionized
partially covering absorber model ($\Delta\chi^2/\Delta\nu=122/9$), and 
gives a good overall representation of the data ($\chi^2_r\sim 1.128$). Spectral
parameters along with their errors are reported in Table~\ref{TABLEFITAVERAGE},
model B. The corresponding theoretical model is plotted in top panel of
Fig.~\ref{FIG6}, while the spectral residuals to the model are plotted in the
bottom panel of Fig.~\ref{FIG6} for each epoch. 

Significant residuals at $E < 1$ keV relative to model B are present only in the
2009 Long Look and Dec. 2004 spectra. The residuals could be either due to an
improper modelization of the absorbers or to an unmodeled component that
emerges in the soft band when the source is in a low flux state. In any case, we
know from the timing analysis performed in Sect.~\ref{timing} that in this
energy range the source shows no spectral variability on short time scales. We
find an improvement in statistics of $\Delta\chi^2/\Delta\nu=42/2$ and
$\Delta\chi^2/\Delta\nu=23/2$ for the addition of two narrow Gaussian emission
lines with centroid energies and normalizations fixed between the epochs. The
best-fit rest frame energies are $E_1^{emi}=0.55^{+0.01}_{-0.01}$ keV and
$E_2^{emi}=0.91^{+0.01}_{-0.01}$ keV, while the equivalent widths $EW_1^{emi}\sim 10-30$ eV
and $EW_2^{emi}\sim 15-40$ eV, depending on the epoch of observation. The 
fit statistics after including the two soft emission lines
is $\chi^2/\nu=1710/1566$. The observed energies might correspond to O VII K and Fe L
emission features, but the exact identification is prohibitive with the pn
spectral resolution. Physically, possible contributors to the soft emission can
be a scattered component accompanied by recombination lines from a photoionized
gas, thermal emission from an underlying starburst component, or a mixture of
the two \citep[e.g. IRAS 13197-1627,][]{2007MNRAS.375..227M}. Unfortunately, PG
1126-041 is not detected by the RGS, and the pn spectral resolution does not
allow the different scenarios to be distinguished, thus we do not try to 
model the soft residuals any further.

As for the high-energy residuals, an excess of counts at
observed energies $E\lesssim 6$ keV is present in the 2009 Long Look data,
suggesting the presence of complexities in the Fe K band. We note that at these
energies PG 1126-041 is highly variable on kilosecond time scales
(Sect.~\ref{timing}). The addition of an Fe I K$\alpha$ emission line, which is forced to
be narrow and has a centroid energy of 6.4 keV in the source rest frame
improves the fit statistics by $\Delta\chi^2/\Delta\nu=15/1$, corresponding to a
significant detection (F-test probability $>99.97$\%). 
The best-fit model statistics are $\chi^2/\nu=1691/1565$, with a reduced
chi square $\chi^2_r=1.080$. There are still shallow residuals in the Fe K band, 
which could be modeled with a broad Gaussian emission line, a relativistic Fe K line,
or a partial covering absorber, in all cases suggesting the possible presence of further
complexities in the iron K band (see M. Giustini Ph.D. thesis for details).
However, given the low statistical significance of the residuals, their appearance only 
in the 2009 Long Look observation, and the high variability of the source at those energies
on a short time scale during the same observation, we do not try to model these residuals
and turn our attention on the time-resolved spectral analysis of the 2009 Long Look data.

\section{Time-resolved spectral analysis}\label{timres}

\begin{figure} 
\begin{center}
\includegraphics[angle=-90,width=84mm]{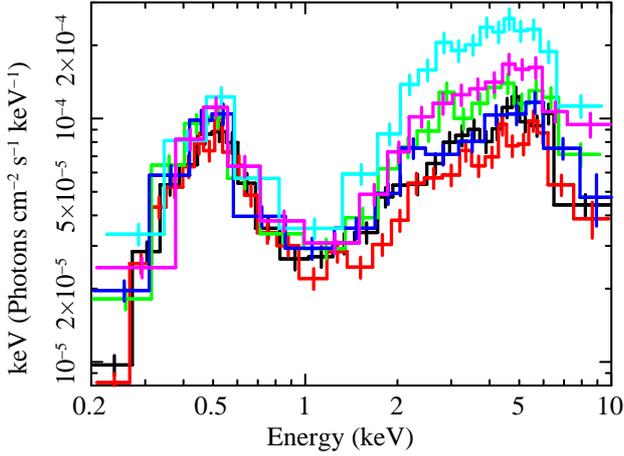} 
\caption{The unfolded 0.2-10 keV EPIC-pn spectra extracted in the different time intervals marked in Fig.~\ref{FIG2}; 
$1^{st}$ (black), $2^{nd}$ (red), $3^{rd}+5^{th}$ (green), $4^{th}+6^{th}$ (blue), $7^{th}$ (light blue), and $8^{th}$ (magenta).
Data points have been visually rebinned to 10 $\sigma$
confidence level.} 
\label{FIG11}
\end{center}
\end{figure}

\begin{table}
\caption{2009 Long Look observation time slices over which time-resolved analysis was performed.\label{tabletimre}}
\centering          
\begin{tabular}{ ccccc}
\hline\hline
Interval & Exposure & Count-rate & $f_{2-10}$ & $f_{0.2-10}$\\
\hline
1 & 21.1 & 0.180$\pm 0.003$ & 0.87$^{+0.08}_{-0.02}$ & 1.00$^{+0.09}_{-0.03}$\\
 2 & 30.0 & 0.156$\pm 0.003$  & 0.71$^{+0.07}_{-0.03}$ & 0.83$^{+0.08}_{-0.03}$\\
 3 & 5.4   & 0.212$\pm 0.007$  & 1.17$^{+0.04}_{-0.05}$ & 1.32$^{+0.04}_{-0.06}$\\
  4 & 7.5   & 0.195$\pm 0.005$  & 0.98$^{+0.06}_{-0.04}$ & 1.12$^{+0.07}_{-0.05}$\\
  5 & 4.6   & 0.215$\pm 0.007$  & 1.17$^{+0.04}_{-0.05}$  & 1.32$^{+0.04}_{-0.06}$\\
6 & 5.4   & 0.194$\pm 0.006$  &  0.98$^{+0.06}_{-0.04}$  & 1.12$^{+0.07}_{-0.05}$\\
 7 & 7.7   & 0.338$\pm 0.007$  & 2.11$^{+0.59}_{-0.08}$ & 2.31$^{+0.62}_{-0.09}$\\
 8 & 9.2   & 0.245$\pm 0.006$  & 1.38$^{+0.14}_{-0.05}$ & 1.54$^{+0.14}_{-0.06}$\\
 \hline
 \end{tabular}
\begin{flushleft}
\tablefoot{Col.(1): Time interval as marked in Fig.~\ref{FIG2}; Col.(2):
Exposure time in ks; Col.(3): 0.2-10 keV pn count rate in ct s$^{-1}$; 
Col.(4): best-fit pn observed flux computed over the 2-10 keV observed band,
in units of 10$^{-12}$ erg cm$^{-2}$ s$^{-1}$;
Col.(5): best-fit pn observed flux computed over the 0.2-10 keV observed band,
in units of 10$^{-12}$ erg cm$^{-2}$ s$^{-1}$.
Errors quoted at 1$\sigma$ confidence level.\\ }
\end{flushleft}
\end{table}
We performed time-resolved spectral analysis on the EPIC pn data extracted in the
eight time intervals of the 2009 Long Look observation marked in the middle
panel of Fig.~\ref{FIG2}. Exposure times and background-corrected count rates are
reported in Table~\ref{tabletimre} for each interval. Given that the
count rates are the same in several intervals, we merged the third and the
fifth intervals together and the fourth and the sixth one together.
The unfolded 0.2-10 keV pn spectrum is plotted in Fig.~\ref{FIG11} for each of the six resulting intervals. 

At energies $E\gtrsim 1.5$ keV there are variations
of about 50\% on a few kiloseconds timescales (e.g., from the
second to the third time interval, and from the seventh to the eighth
time interval), and variations as high as 100\% with respect to the average are
observed during the seventh interval. 
All the spectra were simultaneously fitted to the best-fit model B described in the previous section,
with all the parameters tied between the time intervals except for the power-law normalizations.
The fit to the data is acceptable, $\chi^2/\nu=925/889$, however the model fails to properly reproduce the Fe K band absorption residuals.
Allowing for the highly ionized absorber column density and ionization parameter to vary within
each temporal slice improves the fit by  $\Delta\chi^2/\Delta\nu=45/6$. 
Because of the much lower S/N in individual
temporal slices, it is not possible to assess
whether the absorber varied in column density, ionization state, or a
combination of them. Fluxes computed using this best-fit model are reported in Table~\ref{tabletimre}.

\begin{figure} 
\begin{center}
\includegraphics[width=84mm]{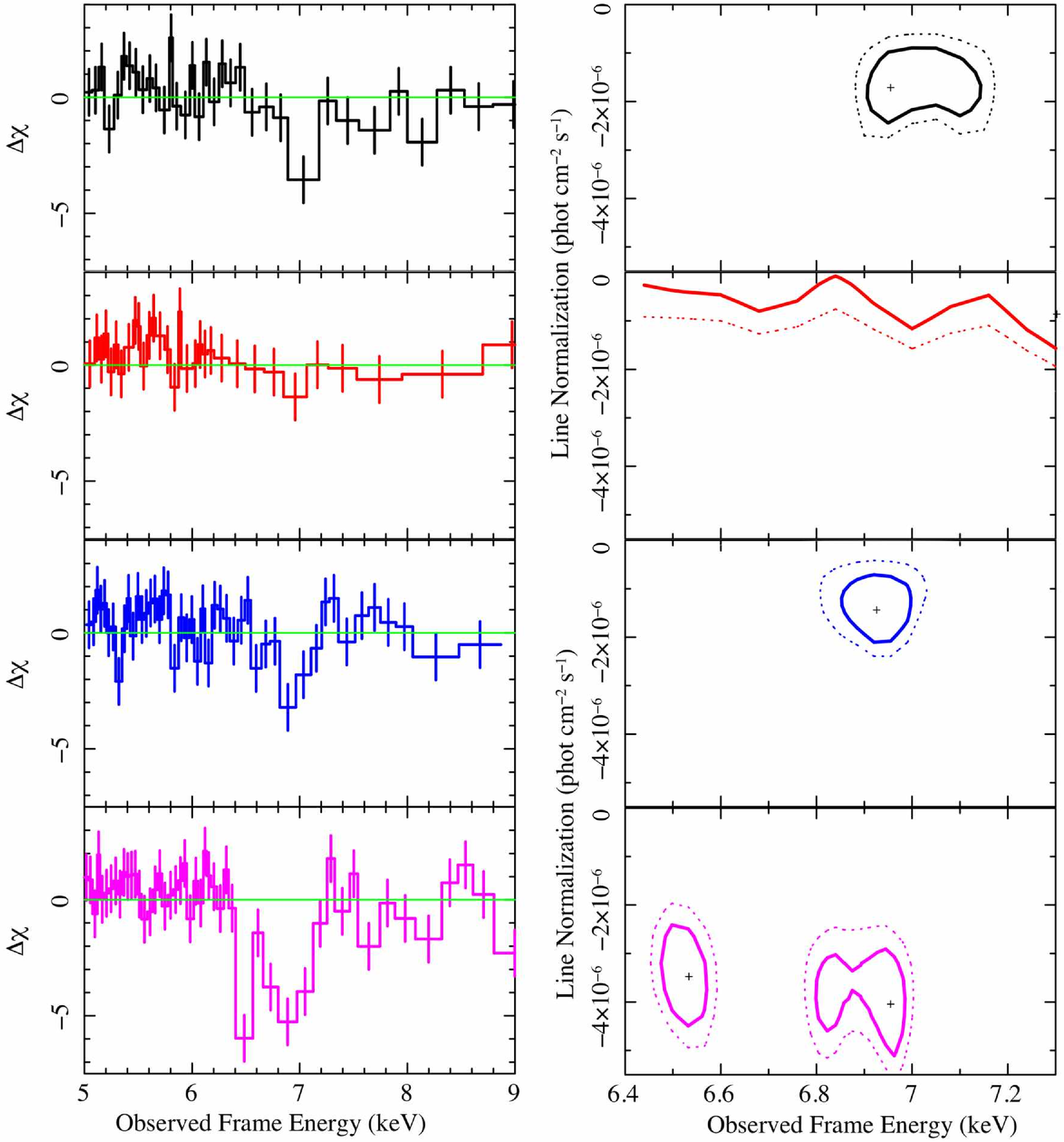} 
\caption{Left panels: 5-9 keV residuals for each temporal slice to model B, 
after the variable highly ionized outflowing absorber was removed. From top to bottom: 1$^{st}$ (black),
2$^{nd}$ (red), 3$^{rd}+4^{th}+5^{th}+6^{th}$ (blue), 7$^{th}+8^{th}$ (magenta) interval.  Right
panels: 68\% (continuous line) and 90\% (dotted line) confidence contours for the energy and intensity of
a narrow Gaussian absorption line used to model the residuals in each slice.\label{FIG12}}
\end{center}\end{figure}

We then removed the highly ionized absorber from the model and visually
inspected the residuals in each temporal slice. Given the similar shape of the
residuals, we merged the third, fourth, fifth, and sixth time intervals together,
and the seventh and eighth time intervals together. The 5-9 keV spectral residuals
are shown in the left column of Fig.~\ref{FIG12} for each of the four resulting
intervals; time is flowing from top to bottom. There is clear variability of the
highly ionized absorber on very short time scales: in particular, there are no
signs of high-energy absorption during the second interval, while the effect
of iron absorption during the last time interval is dramatic. The right column of
Fig.~\ref{FIG12} shows the corresponding 68\% and 90\% confidence contours for
the energy and (negative) normalization of a narrow Gaussian absorption line
used to model the residuals. One can see how the complex absorption features at 
$\sim 7$ keV in the observer frame do disappear 
within the few ks that elapsed between the first and the
second time intervals, and then re-develop during the third to sixth temporal slices.
The nearly $7$ keV absorption trough becomes the deepest and the largest during the seventh-eighth interval,
and its shape becomes quite complex, as confirmed by the confidence contours. 
The intense absorption structure at $\sim 6.5$ keV does develop in a few ks and is also visible in the
merged 2009 Long Look spectra (top panel of Fig.~\ref{FIG5}).
Summarizing, the continuum power-law amplitude oscillations dominate
the observed spectral variability on ks time scales. The highly ionized
absorber is also observed to be variable, while all the other components (i.e.
the photon index, the partial covering moderately ionized absorber, and the
narrow Gaussian emission lines) are found to be constant during the observation.

\begin{figure} 
\centering
\includegraphics[angle=-90,width=75mm]{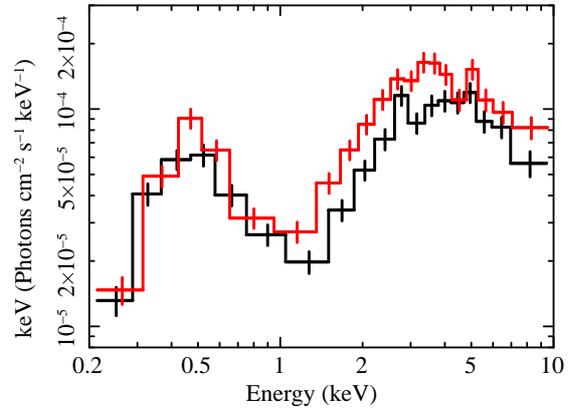} 
\caption{The unfolded 0.2-10 keV EPIC-pn spectra extracted 
during the first 10 ks (black) and 
during the last 10 ks (red) of the Dec. 2004 observation.
Data points have been visually rebinned to 10 $\sigma$ confidence level.\label{2004}}
\end{figure}

As for the other epochs, the Jun. and Dec. 2008 observations were
too short to search for short-term variability. Strong spectral variability was
found instead during the 2004 observation: Fig.~\ref{2004} shows
the unfolded pn spectra extracted during the first 10 ks and during the last 10
ks of the Dec. 2004 observation. The variability pattern is the same as of the 2009
Long Look observation i.e. the source is variable on ks time scales only at energies $E\gtrsim 1$ keV,
and its spectral variabilty is dominated by continuum amplitude variations.

\section{Optical/X-ray photometry\label{OM} }
Thanks to the presence of the OM
aboard XMM-\textit{Newton}, simultaneous optical/X-ray photometry was performed
for each epoch of observation. The OM data were all taken in image mode, and were
reduced and calibrated using standard pipeline procedures through the
\texttt{omichain} SAS task. Background-corrected count rates were converted in
magnitudes for each exposure (exposure times are ranging between 1 and 4 ks). We
were interested in measuring $f_{2500}$ to compute the optical-to-X-ray spectral
index $\alpha_{\rm{ox}}=-0.384\log(\ell_{2500\AA}/\ell_{2\, keV})$, as defined by
\citet{1979ApJ...234L...9T}. We used the weighted mean of the magnitudes
measured in the filters with effective wavelength straddling $2650\, \AA $
in the observed frame ($2500\, \AA$ in the rest frame). The relevant filters are the UVW1 and UVM2
($\lambda_{eff}=2910$ and $2310\, \AA$, respectively). We then corrected
this magnitude for Galactic extinction A$_{2650}=0.382$, estimated using the
extinction law of \citet{1989ApJ...345..245C}. Flux densities for each epoch were then
computed using Vega as a calibrator. Optical photometry results are reported in
Table~\ref{tabOM}. 

\begin{table*}
\caption{Optical photometry\label{tabOM}}
\centering          
\begin{tabular}{ ccccccc}  
\hline\hline
 Epoch       & V                             &  U                           & UVW1 			    & UVM2 			& UVW2 			   & M(2500)  \\
             &  [5430$\,\AA$]  &  [3440$\,\AA$]         &  [2910$\,\AA$]     &  [2310$\,\AA$]     &  [2120$\,\AA$]  &  [2500$\,\AA$] \\
\hline
Dec. 2004 & 14.71$\pm{0.01}$ & 14.09$\pm{0.01}$& 13.87$\pm{0.01}$ & 14.13$\pm{0.02}$& 14.01$\pm{0.04}$ & 13.60   \\
Jun. 2008 & (...) & (...)							 & 13.71$\pm{0.01}$& 13.92$\pm{0.01}$& (...)			    & 13.42  \\
Dec. 2008 & (...)& (...)							& 13.82$\pm{0.01}$ & 13.86$\pm{0.02}$& (...) 			   & 13.46   \\
2009 Long Look & (...)& (...)					& 13.564$\pm{0.004}$& 13.656$\pm{0.008}$& 13.72$\pm{0.02}$& 13.22  \\
\hline
\end{tabular}\\
\begin{flushleft}
\tablefoot{For each filter we list the effective wavelength; the magnitudes measured in each filter are not corrected for
the Galactic extinction, while the extrapolated magnitude at $2500\, \AA$ is corrected for the
Galactic extinction.}
\end{flushleft}
\end{table*}

As for the 2 keV flux density measurements, we used our best-fit model to
measure the Galactic absorption corrected rest frame $f_{2\,keV}$ for each
observation. Flux densities were then converted to luminosity densities, and the
optical-to-X-ray spectral index was computed. It is well known that the
$\alpha_{\rm{ox}}$ spectral index in AGN strongly depends on the intrinsic UV
continuum luminosity
\citep[e.g.,][]{2003AJ....125..433V,2005AJ....130..387S,2006AJ....131.2826S,2010ApJ...708.1388Y}.
We compared the observed $\alpha_{\rm{ox}}$ with the value
$\alpha_{\rm{ox}}(\ell_{2500\AA})$ expected on the basis of the $2500\,\AA$ UV luminosity of
PG 1126-041, using Eq.~2 of \citet{2006AJ....131.2826S}, which gives us the parameter
$\Delta\alpha_{\rm{ox}}\equiv \alpha_{\rm{ox}}-\alpha_{\rm{ox}}(\ell_{2500\AA})$. 
We note that in the UV luminosity range of PG 1126-041 
there is a 3$\sigma$ dispersion in the expected $\alpha_{\rm{ox}}(\ell_{2500\AA})$ of about 0.1, so that only values of 
$|\Delta\alpha_{\rm{ox}}|>0.1$ are statistically significant \citep{2006AJ....131.2826S}.
We also computed the $\ell_{2\,keV}^{corr}$ luminosity densities corrected for intrinsic
X-ray absorption, and the corresponding $\alpha_{\rm{ox}}^{\rm{corr}}$ and
$\Delta\alpha_{\rm{ox}}^{\rm{corr}}$. In Table~\ref{tabAOX} we report
the Optical/X-ray photometry results. 

There are variations in the observed luminosity density of PG 1126-041 both at
UV and at X-ray wavelength. While the observed 2500 $\AA$ flux density
steadily increases from 2004 to 2009, and is varying by about 30\%, the
observed 2 keV flux density increases from 2004 to 2008, and then decreases
again in 2009, with variations as high as a factor of $\sim 4-5$ between the Dec. 2008 and the
2009 Long Look observations. As a result, the observed $\alpha_{\rm{ox}}$ is
variable between the different epochs, spanning from values typical of radio
quiet type 1 AGN (i.e., $\alpha_{\rm{ox}}\sim -1.7$ in the Dec. 2008 observation)
to values typical of Soft X-ray Weak AGN (i.e., $\alpha_{\rm{ox}}\sim -2$ in the
2009 Long Look observation). However, compared to the expected
$\alpha_{\rm{ox}}(\ell_{2500\AA})$, the source is found to be ``ÊX-ray weak'' in all the
epochs, with $\Delta\alpha_{\rm{ox}}\sim -0.3-0.6$. Once the effect of intrinsic
X-ray absorption is taken into account, the maximum observed variations in the 2
keV flux density between the different epochs decrease to about a factor of two.
Consequently, also the $\Delta\alpha_{\rm{ox}}$ decreases to
$\Delta\alpha_{\rm{ox}}^{\rm{corr}}\sim 0.1-0.3$. 
It is worth noting that there could be intrinsic UV absorption at the redshift of the source,
e.g. by dust in the AGN host galaxy; the amount of intrinsic UV absorption is unknown
so we do not try to model it, but we keep in mind that this effect might make the actual (intrinsic) $\alpha_{\rm{ox}}$
 flatter than $\alpha_{\rm{ox}}^{\rm{corr}}$.

\begin{table*}
\caption{Optical/X-ray photometry\label{tabAOX}}
\centering          
\begin{tabular}{ cccccccccccc}  
\hline\hline
 Epoch       &  $\log f_{2500}$ & $\log f_{2 keV}$ & $\log \ell_{2500}$ & $\log \ell_{2\,keV}$ & $\alpha_{\rm{ox}}$ & $\Delta \alpha_{\rm{ox}}$ & $\log \ell_{2\,keV}^{\rm{corr}}$ &$\alpha_{\rm{ox}}^{\rm{corr}}$ & $\Delta \alpha_{\rm{ox}}^{\rm{corr}}$\\
   (1) 	          &(2)                            &(3)                          &(4)     			    & (5)    			&    (6)  			     & (7)                                       &    (8)           & (9) 	  &(10)  \\
\hline
Dec. 2004 &  -25.46& -30.46&  29.49 & 24.49 & -1.92 & -0.52 & 25.28 &  -1.68 & -0.28\\
Jun. 2008 & -25.39 & -30.17& 29.56 & 24.77 & -1.84 & -0.43  & 25.31 & -1.63 & -0.22\\
Dec. 2008 &  -25.40 & -29.80&  29.55 & 25.15 & -1.69 & -0.28 & 25.57& -1.53 & -0.12\\
2009 Long Look &  -25.30& -30.46& 29.64 & 24.49 & -1.98 & -0.56 & 25.26 &  -1.62 & -0.20\\
\hline
\end{tabular}\\
\begin{flushleft}
\tablefoot{Col.(1): Epoch of observation. Col.(2): $2500\, \AA$ rest frame
flux density corrected for Galactic extinction. Col.(3): 2 keV rest frame flux 
density corrected for Galactic absorption. Col.(4): $2500\, \AA$ rest
frame luminosity density corrected for Galactic extinction. Col.(5): 2 keV 
rest frame luminosity density corrected for Galactic
absorption. Col.(6): Observed optical-to-X-ray
spectral index. Col.(7): Difference between the observed $\alpha_{\rm{ox}}$ and
the one expected on the basis of the $\ell_{2500}$ luminosity. Col.(8): 2 keV rest
frame luminosity density corrected for Galactic and intrinsic absorption.
Col.(9): Optical-to-X-ray spectral index corrected for the
intrinsic X-ray absorption. Col.(10): Difference between the corrected
$\alpha_{\rm{ox}}^{\rm{corr}}$ and the one expected on the basis of the
$\ell_{2500}$ luminosity. Flux densities are in [erg s$^{-1}$ cm$^{-2}$ Hz$^{-1}$],
luminosity densities are in [erg s$^{-1}$ Hz$^{-1}$].} 
\end{flushleft} \end{table*}
 
\section{Discussion}\label{DISC}

PG 1126-041 is observed to be highly variable in X-rays on time scales of both
months and kiloseconds. The variability on time scales of months (observed to be
as high as 4$\times$ in flux) is dominated by a spectral component emerging at
$E\lesssim$ 6 keV (Fig. \ref{FIG1}). On the other hand, the RMS variability
analysis performed on the 2009 Long Look pn observation (92 ks of contiguous
good exposure time) revealed the presence of a spectral component emerging only
at $E\gtrsim 1.5$ keV, which dominates the variability on time scales of kiloseconds
(up to 50\% variations in flux, Fig. \ref{FIG2}).

A wealth of information has been obtained both on the intrinsic X-ray continuum
emission of PG 1126-041 and on the reprocessing media that happen to be in the
inner regions of this AGN. 
We discuss our results starting from the inner
regions around the SMBH (i.e., the shortest time scales) and moving farther out
(i.e., the longest time scales). 
We adopt the SMBH mass estimate of $M_{BH}= 1.2
\times 10^{8}\,M_{\odot}$ as given by \citet{2007ApJ...657..102D}. The $M_{BH}$
estimated by \citet{2006ApJ...641..689V} through scaling relations is a factor
of 2.4 lower. Adopting the bolometric luminosity $L_{bol}=10^{12}\,L_{\odot}$
\citep{1989ApJ...347...29S} and assuming an accretion efficiency $\eta=0.1$
gives an accretion rate needed to power PG 1126-041
of $\dot{M}\sim 0.7\,M_{\odot}/$yr and an Eddington ratio $\lambda\equiv L_{bol}/L_{Edd}\sim 0.26$.
The gravitational radius is $r_g\sim 1.8\times 10^{13}$ cm and the corresponding
light travel time is $t_L\sim 600$ s.

Two absorbers are detected in the X-ray spectra of PG 1126-041, a moderately
ionized ($\log\xi^{m.i.}\sim 1.5$ erg cm s$^{-1}$) and a highly ionized
($\log\xi^{h.i.}\sim 3.5$ erg cm s$^{-1}$) one.
Strong variations in both the
intrinsic continuum and of the highly ionized outflowing absorber are responsible
of the observed kilosecond time scale variability. The power-law photon index
$\Gamma\sim 2$ is found not to vary on kilosecond time scales, while the
intensity of the intrinsic power-law spectrum ifollows the pattern of the
1.5-10 keV count rate over the whole observation (see the middle bottom panel of
Fig.~\ref{FIG2} and Table~\ref{tabletimre}), and doubles in a time interval
$\Delta t$ lasting about 8 ks (the seventh interval). The variability time scale of
the intrinsic continuum can set a constraint on the geometrical size $D$ of the
X-ray emission region through the causality argument, $D<c\Delta t\sim 2.4\times
10^{14} $ cm $\sim 13 r_g$. The highly ionized absorber is observed to be
outflowing at $\upsilon_{out}\sim 16,500$ km s$^{-1}$ during the average 2009
Long Look observation. In the time-resolved spectral analysis of the same
exposure, the outflowing absorber is found to be variable over very short time
scales (Fig.~\ref{FIG12}). With the much lower S/N of each temporal slice, it was
not possible to assess whether the absober varied in ionization state, column
density, or blueshift. However, the very short time scale variability suggests
that the absorber is very compact and very close to the X-ray source and that
we are possibly observing rapid mass ejections from the inner regions of the
accretion disk, e.g., the base of an accretion disk wind. We note in particular
how the absorption features do develop in a very short time scale, from the
sixth to the seventh interval, together with the strong continuum flare.

By making some assumptions, we can try to estimate the mass outflow rate
$\dot{M}_{out}$ associated to the highly ionized absorber. 
In general, the mass outflow rate can be written as
$$
\dot{M}_{out}\propto A(r)\,\rho(r)\,\upsilon_{out}(r)
$$
where $A(r)$ is a geometrical factor that accounts for how the flow diverges,
and $\rho(r)$ and $\upsilon_{out}(r)$ are the density and velocity of the flow.
By assuming a spherically symmetric, isotropic, steady flow with a constant velocity, one can write
\begin{equation}
\quad\quad\quad\quad\quad\quad\quad\dot{M}_{out}=4\,\pi\, r^2\,n\,m_{\rm{H}}\, \upsilon_{out}\,C_f\,F_{\rm{V}}
\label{mout}\end{equation}
where $m_{\rm H}$ is the hydrogen atom mass, $r$ the absorber
distance from the central SMBH, $n$  the absorber number density,
$C_f=d\Omega/4\pi$ is the solid angle occupied by the flow as seen by the central point source, 
and  $F_{\rm V}$ the volume filling factor. 

If we use appropriate values for PG 1126-041, we can rearrange the expression and write
$$
\dot{M}_{out}=0.33\,\left(n_{11} \right)\,\left( r_{10} \right)^2 \,\left( \frac{\upsilon_{out}}{16,500\,\rm{km\,s}^{-1}} \right) \,\left( \frac{C_f}{0.2} \right)\,F_{\rm{V}}\:M_{\odot}\,\rm{yr}^{-1}
$$
where $n_{11}=n/10^{11}$ cm$^{-3}$, $r_{10}=r/10r_g$. In the expression above,
the geometrical covering factor $0<C_f<1$ and volume filling factor
$F_{\rm{V}}\leq 1$ are unknown, and depend on the wind geometry and duty cycle.
The density and the radial distance of the flow are also unknown. 

Assuming that
the outflowing absorber is a thin spherical shell of gas in photoionization
equilibrium, one can use the definition of the ionization parameter $\xi\equiv
L_{ion}/n r^2$ to replace the unknown density and distance with the ratio
$L_{ion}/\xi$. However, in addition to the strong assumptions made regarding the spherical geometry
of the outflow, there are uncertainties in the ionizing
continuum luminosity, and especially in the actual value of the ionization
parameter that may affect our mass outflow rate estimates. 

\begin{figure} 
\centering
\includegraphics[width=74mm]{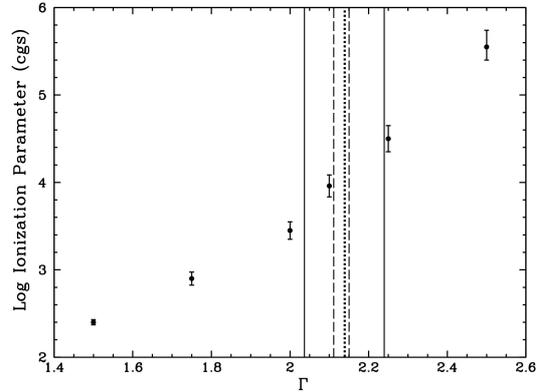} 
\caption{Measured $\xi$ for the highly ionized outflowing absorber in the 2009 Long Look
spectra, with different $\Gamma_{ion}$ used as input in the \texttt{Xstar} simulations;
Error bars are purely statistical errors at 1$\sigma$ confidence level.
The vertical dotted line marks the best-fit measured $\Gamma$ for the 2009 Long Look spectra, along with its
statistical (dashed lines) and systematic (continuous line) errors that have been estimated considering
the uncertainties of the power-law normalization and of the ionized absorbers.}\label{FIGZ}
\end{figure}

In our spectral analysis results we quoted only the statistical
errors on $\xi$: the systematic errors related to the input parameters used in
the photoionization code (\texttt{Xstar} in our case) can be much larger,
especially for highly ionized absorbers where only a few spectral transitions
are available (e.g., Fe XXV, Fe XXVI). The major dependence of the best-fit $\xi$ value is related to
the slope $\Gamma_{ion}$ adopted for the ionizing power-law continuum. This is
because $\xi\propto n_{\gamma}/n_{\rm{H}}$, where $n_{\gamma}$ is the ionizing
photons number density, $n_{\rm{H}}$ the photoionized gas number density. By fixing the
ionizing luminosity, a steeper $\Gamma_{ion}$ results in a smaller number of
high-energy photons able to strip electrons from highly ionized Fe than for a
flatter $\Gamma_{ion}$. This leads to higher best-fit values measured for $\xi$
when using a steeper ionizing continuum. To show this dependence, we generated
several \texttt{Xstar} absorber models using the same input for all the physical
parameters, but different slopes $\Gamma_{ion}$. We then
fitted the 2009 Long Look average spectra to each of these models, recording
the best-fit $\xi$ value measured for the highly ionized absorber responsible
for the absorption features visible in Fig.~\ref{FIG5}. Results are plotted in
Fig.~\ref{FIGZ}: the systematic error on $\xi$ is very large, and even 
restricting the slope of the ionizing continuum to values typical of type 1 AGN, $1.8 < \Gamma_{ion} < 2.2$,
gives a systematic uncertainty on $\xi$ of a factor of about 30. 
Using the measured slope $\Gamma$ for PG 1126-041 along with its systematic error gives an
uncertainty on our measured $\xi^{h.i.}$ of a factor of $\sim 6$. 

Just as for a comparison with other literature
results, we can roughly estimate the kinetic luminosity $L_{kin}$ associated to the PG 1126-041 outflow
as follows.  We take the best estimate for $\xi^{h.i.}$ from Fig.~\ref{FIGZ}, considering the
best-fit measured photon index $\Gamma=2.15$ as the slope of the ionizing continuum. For the
ionizing luminosity, we take the measured unabsorbed 2-10 keV luminosity listed in Table~\ref{properties}.
Substituting the ratio of these two quantities to the product $r^2 n$ in Eq.~\ref{mout},
and using the best-fit outflow velocity, one can write
$$
L_{kin}=\frac{1}{2}\dot{M}_{out}\upsilon_{out}^2\sim 3\left(\frac{C_f}{0.2}\right)\left(\frac{F_V}{1.0}\right)\times 10^{43}\,\rm{erg\: s^{-1}}\rm{,}
$$
i.e., less than one hundreth of the bolometric luminosity of PG 1126-041. 

The uncertainties on the physical quantities related to the outflow are
huge, and the assumptions made to derive the expression for the mass outflow
rate and kinetic luminosity are likely to be incorrect. The presence of the
accretion disk breaks the spherical symmetry by itself; furthermore, the highly
ionized absorber is observed to be variable on very short time scales: the
geometrical and dynamical effects are most probably very important in accretion
disk wind scenarios. For these reasons, it is very difficult to give a
meaningful (i.e. with less than two order of magnitudes uncertainties)
estimate of the mass outflow rate and of its kinetic efficiency. Although one
can not exactly assess the impact of the fast outflow of PG 1126- 041 on its
environment, we stress that its detection is statistically solid and
model-independent.

Turning our attention to the larger scales, we note that at $E\lesssim 1.5$ keV
the power-law variability is smeared out by the presence of a soft constant
component, and at $E < 1$ keV the highly ionized absorber opacity drops (see top
panel of Fig.~\ref{FIG6}), so explaining the constant soft flux measured during
the whole 2009 Long Look observation. As for the soft component, we modeled it
as the direct continuum emission that escapes unabsorbed by a partially covering
ionized gas. However, the low fraction of unabsorbed flux, $(1-C_f)\sim 0.03$,
could also physically correspond to a scattered component, that would be
naturally associated to the presence of ionized gas along the line of sight.
With the moderate pn spectral resolution it is not possible to distinguish among
the different scenarios, and unfortunately PG 1126-041 is not detected by the
RGS. However, we stress that the adopted modeling for the soft band constant
component influences neither the absorbers nor the continuum physical
parameters.

The PG 1126-041 X-ray variability on long time scales is dramatic, and is
especially pronounced at energies $E\lesssim 6$ keV. The $f_{\rm{2\,keV}}$ is
observed to be variable up to 4.5 higher in the six months elapsed between the Dec.
2008 and the 2009 Long Look observations (Fig.~\ref{FIG1}). The moderately ionized absorber is
detected in every \textit{XMM-Newton} observation of PG 1126-041. Its
column density varies significantly between the minimum $N_W^{m.i.}\sim 3\times 10^{22}$
cm$^{-2}$ measured in the Dec. 2008 observation and the maximum $N_W^{m.i.}\sim
1.4\times 10^{23}$ cm$^{-2}$ measured in the 2009 Long Look
observations. 
In fact, the moderately ionized absorber column density variations
dominate the spectral variability observed on month time scales, together with
amplitude variations of the intrinsic continuum up to a factor of two (Table \ref{TABLEFITAVERAGE}).
The ionization state of the moderately ionized absorber $\log\xi^{m.i.}\sim 1.3-1.7$ 
is typical of the X-ray warm absorbers, however the measured column density are much higher
than what is usually observed in warm absorbers \citep[see e.g.][]{2007MNRAS.379.1359M}.

A possible physical scenario is that of a UV line-driven accretion disk wind like the one
developed by \citet{2004ApJ...616..688P} and modeled by
\citet{2010MNRAS.408.1396S}. These authors computed the simulated X-ray spectra
for a nonspherical, hydrodynamical accretion disk wind in an AGN with
$M_{BH}=10^8M_{\odot}$, accreting with $\lambda=L/L_{Edd}=0.5$, i.e., with
conditions that are not too different to those observed in PG 1126-041. The simulated
spectra change significantly for different inclination angles and different
snapshots of the flow, confirming both the nonspherical character and the
highly dynamical behavior of the X-ray spectrum associated with such winds. The
spectral signature of such flows includes both absorption and emission features.
For example, the iron K band is shaped both by absorption by a highly ionized
(Fe XXV, Fe XXVI) outflowing phase of the wind, and by emission and scattering
off the highly ionized base of the flow, which produces a Compton bump that
mimics a red-skewed iron K line. In the soft band ($E<1$ keV), there are
contributions by a blend of emission lines from elements such as C, O, Fe, and
bremsstrahlung emission by the wind. 

From a dynamical point of
view, the same model also predicts a flow component that qualitatively shows the
same behavior of the highly ionized absorber. At intermediate inclination angles
above the disk plane and above the ``skin'' of the fast wind, where the gas is
most exposed to the X-ray ionizing source, recurrent instabilities with blobs
(or ``puffs'') of high density can develop. Some of these puffs quickly become
overionized and fail to become part of the wind, while others can be able to be
efficiently accelerated to velocities even higher than those of the fast
wind and well above the local escape velocity. This flow component would be
observationally detectable in the X-ray spectra via large column density variations
and sporadic high-velocity ejections, on short timescales, similar to what was
observed during the 2009 Long Look observation of PG 1126-041. This may suggest that
radiation-driven accretion disk wind models can account in a self-consistent way
for most of the observed X-ray spectral features in the mini-BAL QSO PG
1126-041. Models for the X-ray spectra of accretion disk winds that can be
fitted to real data are still under developement, therefore we must confine ourselves
to qualitative considerations. However, we
note that in this scenario the radial flow assumption is invalid, and the
mass outflow rate associated with the highly ionized phase is overall very
different than in the spherical case.

The moderately ionized absorber, on the other hand, is not variable on short time scales 
but is strongly variable over time scales of months.
Its observed variability is  consistent with the scenario
suggested to explain the variability in the UV absorption lines of the 
mini-BAL QSO HS 1603+3820 over rest frame time scales of weeks and months
\citep{2007ApJ...660..152M, 2010ApJ...719.1890M}. In this scenario the variability
of the UV absorption lines is the result of fluctuations in the continuum, which is 
caused by a screen of clumpy, highly-ionized gas between the UV absorber and the continuum source.
This clumpy screen has similar properties as the moderately ionized absorber we have
found in PG 1126-041.

The observed optical-to-X-ray spectral index is found to be highly variable,
following the moderately ionized absorber variability. Although $\alpha_{ox}$
never gets $<-2$, when compared with the expected $\alpha_{ox}(\ell_{2500\AA})$
based on the UV continuum luminosity, it is found that PG 1126-041 is observed to
be ``Êsoft X-ray weak'', $\Delta\alpha_{ox}=-(0.3-0.6)$ depending on the epoch of
observation. After correcting for the intrinsic X-ray absorption, the source is still
slightly X-ray weak, $\Delta\alpha_{ox}^{corr}=-(0.1-0.3)$. 
The observed $\alpha_{\rm{ox}}$ variability driven by X-ray absorption variability is similar to what \citet{2008A&A...483..137B} observed
in the mini-BAL QSO PG 1535+547.
The 2-10 keV
unabsorbed luminosity is still very low, $\langle L_{2-10}\rangle\sim 2\times
10^{43}$ erg s$^{-1}$, and compared with $L_{bol}$ gives a very high bolometric
correction, $\kappa_{bol}\sim 200$, that is however still compatible with the
$\kappa_{bol}-\alpha_{ox}$ relation as in \citet{2010A&A...512A..34L}.

\section{Conclusions}\label{CONC}

Our XMM-\textit{Newton} observational campaign on the mini-BAL QSO PG 1126-041
allowed us to characterize the complex high-energy spectral behavior of the
source with unprecedented sensitivity. We analyzed data for four different
pointed observations that in all span 4.5 yr.

The most evident and model-independent result is the detection of high column
densities of ionized gas along the line of sight, which contribute to the
observed spectral variability of the highly ionized
and moderately ionized phase  on kilosecond and month time scales, respectively.
In particular, the highly ionized absorber is found to be variable on very short (a few ks)
time scales, suggesting an origin very close to the SMBH.
Also the intrinsic X-ray continuum emission is observed to be variable
in intensity, on both short and long time scales. 

Overall our findings are qualitatively consistent with radiation-driven,
accretion disk wind model predictions, where one expects a high column density
of X-ray absorbing gas shielding the portion of the wind that is accelerated by UV photons.
The hot X-ray component of the wind is
also expected to be highly variable with time in such a model. The mini-BAL appearance of the
wind of PG~1126-041 is qualitatively consistent with an orientation effect, in
which our line of sight and the plane of the accretion disk make a larger angle
than when viewing classical BAL QSOs. We are looking at a narrower range of UV
velocities through the wind (hence the small width of the UV absorption
features) and at the transition zone between the fast UV wind and the hot X-ray
component of the flow, where we expect strong X-ray spectral variability.

The present X-ray observational campaign of the mini-BAL QSO PG 1126-041 has
demonstrated that long and short term variability studies are quite powerful tools
for identifying the physical mechanisms at work and for mapping the dynamics of the
inner accretion/ejection flow in AGN.

\section*{Acknowledgments}
MG, MC, MD, and CV acknowledge financial support from the ASI/INAF contracts I/088/06/0
and I/009/10/0.
GC and MG acknowledge support provided by NASA grant NNXlOAEllG.
DP and MG  acknowledge support provided by the Chandra award
TM0-11010X issued by the Chandra X-ray Observatory Center, which is
operated by the Smithsonian Astrophysical Observatory for and on behalf of
NASA under contract NAS 8-39073. ME acknowledges support from the National 
Science Fundation under grant AST-0807993.
GP acknowledges support via an EU Marie Curie Intra-European 
Fellowship under contract no. FP7-PEOPLE-2009-IEF-254279. We thank the referee for the thoughtful comments that helped improve
the article presentation. This research has made use of the NASA/IPAC Extragalactic Database (NED) which
 is operated by the Jet Propulsion Laboratory, California Institute of
 Technology, under contract with the National Aeronautics and Space
 Administration. 
 
 \bibliographystyle{aa} \bibliography{bibMB}

\label{lastpage}

\end{document}